\documentclass[aps, pra, twocolumn, superscriptaddress, amsmath,  tightenlines, longbibliography]{revtex4-1}

\usepackage{dcolumn}
\usepackage{graphicx}
\usepackage{mathrsfs}
\usepackage{subfigure}
\usepackage{booktabs}
\usepackage{amsmath}
\usepackage{physics}
\usepackage{dsfont}
\usepackage{amstext}
\usepackage{amssymb}
\usepackage{amsbsy}
\usepackage{bbm}
\usepackage{amsthm}
\usepackage{graphicx}
\usepackage{color}
\usepackage[colorlinks,citecolor=blue]{hyperref}

\usepackage{url}
\usepackage[colorlinks]{hyperref}
\hypersetup{%
	plainpages=true,
	breaklinks=true,       
	hypertexnames=false,  
	pageanchor=true,
	colorlinks=true,
	linkcolor={blue},
	citecolor={red},
	urlcolor={blue},
	anchorcolor={black}
}

 \makeatletter

\newcommand{\Rmnum}[1]{\expandafter\@slowromancap\romannumeral #1@}
\makeatother

\begin{document}
\title{Sudden death of entanglement with Hamiltonian ensemble assisted by auxiliary qubits}
\author{Congwei Lu}
\thanks{These authors contributed equally to this work}
\affiliation{Department of Physics, Applied Optics Beijing Area Major Laboratory, Beijing Normal University, Beijing 100875, China}
\author{Wanting He}
\thanks{These authors contributed equally to this work}
\affiliation{Department of Physics, Applied Optics Beijing Area Major Laboratory, Beijing Normal University, Beijing 100875, China}
\author{Jun Wang}
\thanks{These authors contributed equally to this work}
\affiliation{Department of Physics, Applied Optics Beijing Area Major Laboratory, Beijing Normal University, Beijing 100875, China}
\author{Haibo Wang}
\affiliation{Department of Physics, Applied Optics Beijing Area Major Laboratory, Beijing Normal University, Beijing 100875, China}
\author{Qing Ai}
\email[E-mail: ]{aiqing@bnu.edu.cn}
\affiliation{Department of Physics, Applied Optics Beijing Area Major Laboratory, Beijing Normal University, Beijing 100875, China}

\date{{\small \today}}


\begin{abstract}
In this paper, we theoretically propose a method to simulate the longitudinal relaxation of a single qubit by coupling it to an auxiliary qubit. In order to mimic the finite-temperature relaxation, we utilize the Hamiltonian-ensemble approach [\href{http://dx.doi.org/10.1103/PhysRevX.6.031023}{Kropf, Gneiting, and Buchleitner, Phys. Rev. X \textbf{6}, 031023 (2016)}]. The longitudinal relaxation arises as a consequence of the ensemble average and the interaction between the working qubit and the auxiliary qubit. Furthermore, we apply this approach to investigate the influence of the longitudinal relaxation and the transverse relaxation on the entanglement dynamics of two qubits. It is discovered that the sudden death of the entanglement will occur as long as the longitudinal relaxation is present. The transverse relaxation assists the longitudinal relaxation and thus accelerates the finite-time disentanglement.
\end{abstract}
\maketitle

\section{Introduction}
\label{sec:introduction}

In an open quantum system, the interaction between the system and the environment will lead to the exchange of information and energy between them. As a result, the dynamic behavior of the system is quite different from that of an isolated system \cite{Leggett1987RMP,Breuer2002,Zhang2012PRL,Breuer2016RMP,deVega2017RMP,Chen2022NPJQI}. Generally, there are two types of relaxations, i.e., transverse relaxation and longitudinal relaxation. The latter will result in population transfer and decay of the off-diagonal terms of the density matrix, while the former will only decrease the coherence. These two behaviors play a crucial role in quantum information processing.

Recently, it was proposed that the quantum dynamics of an open quantum system can be simulated by the ensemble-averaged state of many random isolated systems \cite{Buluta2009S,Georgescu2014RMP,Kropf2016PRX,Gneiting2017PRL,Chen2018PRL,Wang2018NPJQI,Zhang2021FP}. The process of ensemble averaging over each random realization will result in averaging all random phases, thus inducing the loss of phase information, i.e., dephasing. However, due to the classical property of noise, most of the previous quantum simulation approaches can only simulate the longitudinal relaxation at the high-temperature limit \cite{Soare2014NP,Wang2018NPJQI,Zhang2021FP,Soare2014PRA,Zhen2016PRA}. The quantum simulation of the longitudinal relaxation at finite temperature is rarely studied. Since the dissipative environment can cause finite-time disentanglement \cite{Yu2004PRL,Almeida2007S,Yu2009S}, enhanced relaxation at avoided level-crossing \cite{Yang2020AP,Onizhuk2021PRXQ,Head-Marsden2021CR}, and optical non-reciprocity by detailed balance \cite{Yao2022A}, it may be interesting to in depth understand the influence of transverse relaxation and longitudinal relaxation on the sudden death of entanglement and its dynamic simulation.

In this paper, in order to simulate the longitudinal relaxation at finite temperature, we introduce an auxiliary qubit which interacts with the working qubit. In order to mimic the longitudinal relaxation, we effectively prepare a large number of systems including the auxiliary qubit and the working qubit. They evolve from the same initial state and in each realization the interaction strength is subject to a Gaussian random distribution. By averaging over the different realizations, we can effectively simulate the longitudinal relaxation. Our analytical results demonstrate that this approach can well simulate the finite-temperature longitudinal relaxation with the dissipation rate linearly dependent on time. Here, the dissipation rate scales linearly with the variance of the random interaction strength between the auxiliary qubit and the working qubit. The initial state of the auxiliary qubit determines the distribution of the steady state.

We further investigate the effects of longitudinal and transverse relaxation on the entanglement of two qubits. We let the first working qubit interact with the auxiliary qubit to mimic the longitudinal relaxation, and apply a random field on the second working qubit to simulate the transverse relaxation. These two working qubits are initialized in the maximum-entangled state and the concurrence is utilized to characterize the dynamics of entanglement. Our simulations show that due to the longitudinal relaxation, the sudden death of entanglement happens at a finite time. In this case, the transverse relaxation of the second working qubit will accelerate the finite-time disentanglement of the two qubits. However, if the longitudinal relaxation is absent, the entanglement will go to zero when the time approaches infinity.

The rest of the paper is structured as follows. In Sec.~\ref{sec:model}, we introduce the quantum-simulation approach by Hamiltonian ensemble. In Sec.\ref{sec:SingleQubit}, we simulate the longitudinal relaxation of a single qubit at a finite temperature. The effects of the noise fluctuation and interaction strength on entanglement sudden death are investigated in Sec.~\ref{sec:disentanglement}. Finally, we conclude our main discoveries in Sec.~\ref{sec:conclusion}.

\begin{figure}[!tb]
	\centering
	\includegraphics[width=8.4cm]{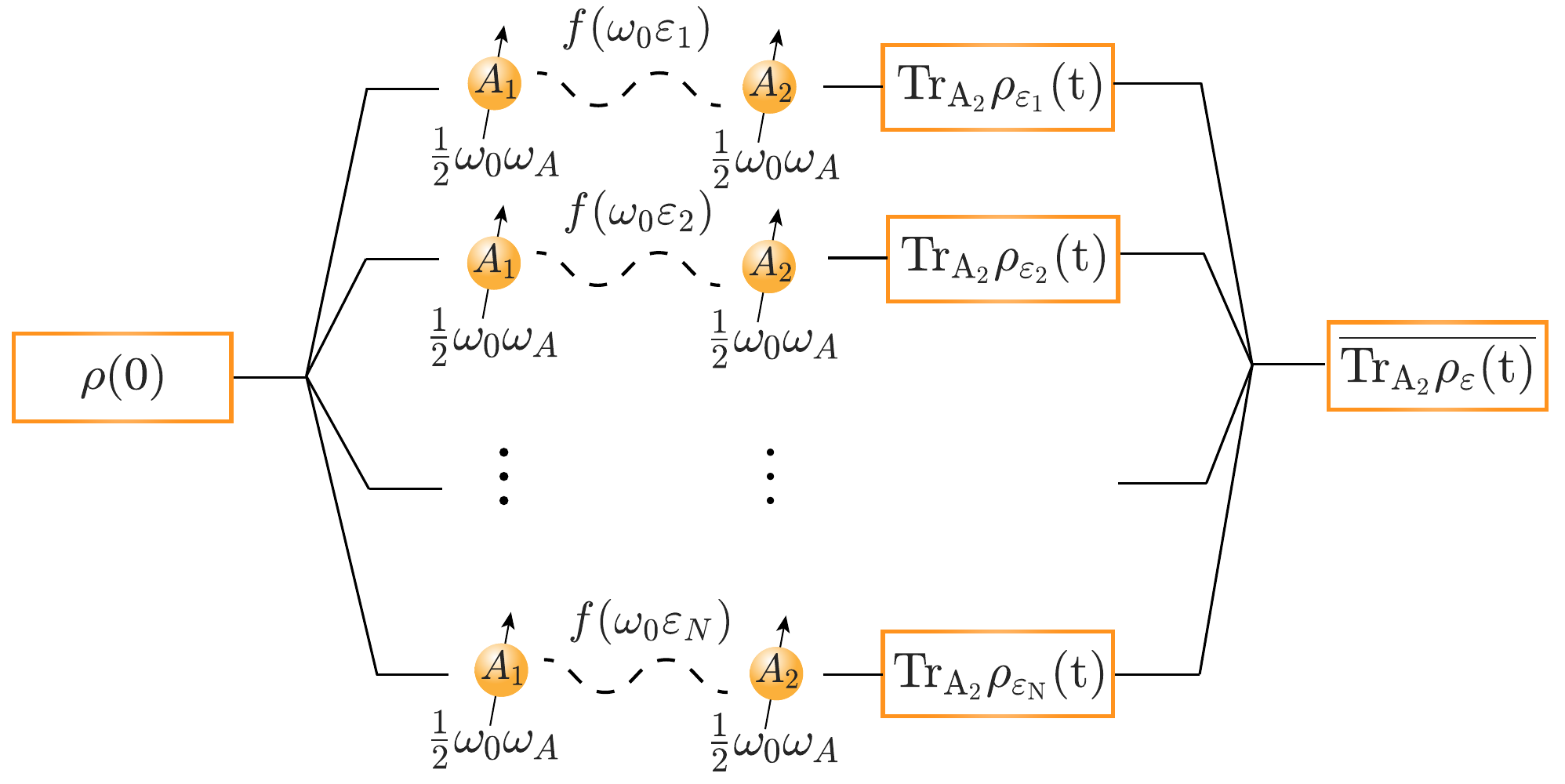}
	\caption{Schematic diagram for simulating the longitudinal relaxation of a single qubit by the Hamiltonian-ensemble approach assisted by an auxiliary qubit. Each realization is composed of a working qubit ${A_1}$ and an auxiliary qubit ${A_2}$. In each realization, they evolve independently from the same initial state $\rho(0)$. The ensemble-averaged state $\overline{\operatorname{Tr}_{A_2}\rho_{\varepsilon}(t)}$ is then obtained by averaging over all reduced density matrix $\operatorname{Tr}_{A_2}\rho_{\varepsilon_{i}}(t)$ ($i=1,2,\cdots,N$) of the working qubit. } 	
\label{fig:SingleQubitSketch}
\end{figure}

\section{Hamiltonian-Ensemble Approach}
\label{sec:model}

First of all, we shall give a brief introduction to the quantum-simulation approach by an ensemble of Hamiltonians \cite{Kropf2016PRX,Chen2018PRL,Wang2018NPJQI,Zhang2021FP}, as schematically illustrated in Fig.~\ref{fig:SingleQubitSketch}. A general open quantum system can be characterized by a total Hamiltonian $\hat{H}_{T}=\hat{H}_{S}+\hat{H}_{E}+\hat{H}_{I}$ \cite{Breuer2002}, where $\hat{H}_{S}$ is the system Hamiltonian, $\hat{H}_{E}$ is the environment Hamiltonian, and $\hat{H}_{I}$ represents their interaction. The time evolution of the open system can be described as $\rho_T(t)=\hat{U} \rho_{T}(0) \hat{U}^{\dagger}$, with $\hat{U}=\exp (-i \hat{H}_T t / \hbar)$. Thus, the density matrix of the system can be obtained by partially tracing over the environmental degrees of freedom, i.e., $\rho_{S}(t)=\operatorname{Tr}_E[\rho_T(t)]$. To simulate the open quantum dynamics, we utilize the Hamiltonian ensemble
\begin{align}\label{Eq1}
\{(\hat{H}_\varepsilon, p_\varepsilon)\},
\end{align}
where the subscript $\varepsilon$ denotes each realization in the ensemble.  The single realization Hamiltonian $\hat{H}_\varepsilon$ occurring with probability $p_\varepsilon$ reads
\begin{align}\label{Eq2}
\hat{H}_\varepsilon=\hat{H}_S+\hat{H}_{E}^{\varepsilon}+\hat{V}_{\varepsilon}.
\end{align}
$\hat{H}_{E}^{\varepsilon}$ and $\hat{V}_{\varepsilon}$ are utilized to simulate the environment and its interaction with system. We suppose that each realization begins from the same initial state $\rho_\varepsilon(0)=\rho(0)$. The corresponding evolution at time $t$ is given by $\rho_\varepsilon(t)=\hat{U}_\varepsilon \rho(0) \hat{U}_{\varepsilon}^{\dagger}$, with $\hat{U}_\varepsilon=\exp (-i \hat{H}_\varepsilon t / \hbar)$. Finally, we trace over the environmental degree of freedom in each realization and then average over all realizations, i.e.,
\begin{align}\label{Eq3}
\langle\rho(t)\rangle&=\overline{\operatorname{Tr}_E\rho_\varepsilon(t)}=\int d\varepsilon p_\varepsilon
 \operatorname{Tr}_E\rho_\varepsilon(t).
\end{align}
Hereafter, all ensemble-averaged quantities will be marked with a bar. In the next section, as an example, we utilize the ensemble-averaged quantum dynamics of the state $\langle\rho(t)\rangle$ to simulate the longitudinal relaxation behavior of a single qubit.

\section{Longitudinal Relaxation of A Single Qubit}
\label{sec:SingleQubit}

In the previous investigations, due to the classical property of the noise, the quantum simulation approach can only simulate the longitudinal relaxation at the high-temperature limit \cite{Wang2018NPJQI,Zhang2021FP,Chen2022NPJQI}. In this paper, we introduce an auxiliary qubit, which interacts with the working qubit as schematically illustrated in Fig.~\ref{fig:SingleQubitSketch}, in order to simulate the longitudinal relaxation at finite temperatures. In the Hamiltonian ensemble, the Hamiltonian of a realization reads
{\begin{align}\label{Eq4}
\hat{H}_\varepsilon=&\frac{\omega_0}{2}(\omega_A\sigma_{z}^{A_1}+\omega_A\sigma_{z}^{A_2})+f(\omega_0\varepsilon)\sigma_{+}^{A_1}\sigma_{-}^{A_2}\nonumber\\
&+f^\ast(\omega_0\varepsilon)\sigma_{-}^{A_1}\sigma_{+}^{A_2},
\end{align}}
where we set $\hbar=1$, $\sigma_z$ is the Pauli matrix,  $\sigma_{\pm}$ are the raising and lowering operators and $\omega_0$ is the unit for frequency. Hereafter, we set $\omega_0=1$ in the following simulations.  $f(\varepsilon)=f^{\ast}(\varepsilon)$ is the coupling strength between the working qubit and the auxiliary qubit. For simplicity, the two-qubit product states are relabeled as
\begin{equation}\label{Eq5}
\begin{split}
|1\rangle_{A_1 A_2}=|+-\rangle_{A_1 A_2},~|2\rangle_{A_1 A_2}=|-+\rangle_{A_1 A_2}, \\
|3\rangle_{A_1 A_2}=|++\rangle_{A_1 A_2},~|4\rangle_{A_1 A_2}=|--\rangle_{A_1 A_2},
\end{split}
\end{equation}
where $|\pm\pm\rangle_{A_1 A_2}\equiv|\pm\rangle_{A_1} \otimes|\pm\rangle_{A_2}$ denote the eigenstates of Pauli operator $\sigma_z^{A_1} \otimes \sigma_z^{A_2}$. Since $\sigma_z^{A_1} \otimes \sigma_z^{A_2}$ is the conserved quantity, i.e., $[\sigma_z^{A_1} \otimes \sigma_z^{A_2},~\hat{H}_\varepsilon]=0$, we can rewrite $\hat{H}_\varepsilon$ as a block-diagonal matrix in the basis listed in Eq.~(\ref{Eq5}),
\begin{align}\label{Eq6}
\hat{H}_\varepsilon=\left(\begin{array}{cc}
\hat{H}_\varepsilon^{-}& 0 \\
0 & \hat{H}_\varepsilon^{+}
\end{array}\right),
\end{align}
where
{\begin{equation}\label{Eq7}
\begin{split}
\hat{H}_\varepsilon^-&=f(\varepsilon)\sigma_x, \\
\hat{H}_\varepsilon^+&=\omega_A\sigma_z.
\end{split}
\end{equation}}
The total evolution operator $\exp(-i\hat{H}_\varepsilon t)$ can be represented as
\begin{align}\label{Eq8}
\hat{U}_\varepsilon=\left(\begin{array}{cc}
\hat{U}_\varepsilon^{-}& 0 \\
0 & \hat{U}_\varepsilon^{+}
\end{array}\right),
\end{align}
where
{\begin{equation}\label{Eq9}
\begin{split}
\hat{U}_\varepsilon^-&=\cos[f(\varepsilon)t]-i\sin[f(\varepsilon)]\sigma_x,\\
\hat{U}_\varepsilon^+&=\cos(\omega_At)-i\sin(\omega_At)\sigma_z,
\end{split}
\end{equation}}

\begin{figure*}[!tb]	
	\centering
	\includegraphics[width=18cm]{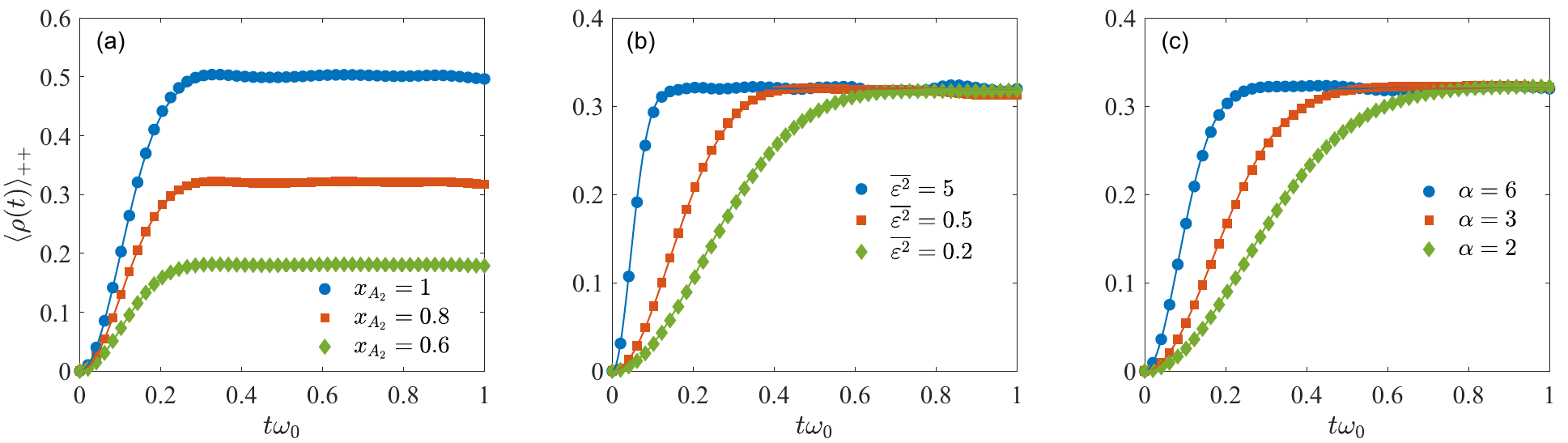}
	\caption{The longitudinal relaxation against $x_{A_2}$, $\overline{\varepsilon^2}$ and $\alpha$ for $s=0$. The population of the subsystem ${A_1}$ at $|+\rangle$ under ensemble average (a) for $x_{A_2}=1,~0.8,~0.6$, when $\overline{\varepsilon^2}=1$ and $\alpha=5$, (b) for $\overline{\varepsilon^2}=5, 0.5, 0.2$ when $x_{A_2}=0.8$ and $\alpha=5$, (c) for $\alpha=6, 3, 2$, when $x_{A_2}=0.8$ and $\overline{\varepsilon^2}=1$. The analytic results are shown as solid lines, while the dotted lines are generated by $N=8000$ random samples with $\overline{\varepsilon}=0$. }  \label{figure1}
\end{figure*}

We assume that the initial state is a product state $\ket{\psi(0)}=\ket{\phi(0)}_{A_1}\otimes\ket{\varphi(0)}_{A_2}$, where the working qubit ${A_1}$ is in the state $\ket{\phi(0)}_{A_1}=\ket{-}_{A_1}$, and the auxiliary qubit ${A_2}$ is in a superposition state $\ket{\varphi(0)}_{A_2}=x_{A_2}\ket{+}_{A_2}+y_{A_2}\ket{-}_{A_2}$. The reduced density matrix of the working qubit ${A_1}$ can be obtained by partially tracing over ${A_2}$,
\begin{align}\label{Eq10}
\rho_{A_1}(t)=\operatorname{Tr}_{A_2}(\hat{U}_\varepsilon\ket{\psi(0)}\bra{\psi(0)}\hat{U}_\varepsilon^\dagger).
\end{align}
For simplicity, we assume that the coupling strength is of the form as
\begin{eqnarray}\label{Eq11}
f(\varepsilon)=\alpha(\varepsilon-s),
\end{eqnarray}
where $\varepsilon$ is a random number for each realization. The physical meanings of parameters $\alpha$ and $s$ depend on different physical implementations. Here, we propose a solution that can implement our model. The two-qubit interaction in Eq.~(\ref{Eq4}) can be realized by the effective dipole-dipole interaction between two two-level atoms coupled to the photonic crystal modes, where the atomic resonance is close to one of the band edges of the photonic crystal \cite{Douglas2015NPhoton}. The effective interaction between the two atoms in the interaction picture is
\begin{eqnarray}\label{Eq12}
	H_{\text{I}}\approx \frac{g^2_c}{2\omega_A}F(z_{A_1},z_{A_2})(\sigma_{+}^{A_1} \sigma_{-}^{A_2}+\sigma_{-}^{A_1} \sigma_{+}^{A_2}),
\end{eqnarray}
where $g_c=g\sqrt{2\pi/L}$, $g$ is the coupling strength between atom and photon, and $L$ is the length scale of the photon decays from the atomic position $z_j$. The tunable function $F(z_{A_1}, z_{A_2})$ decays exponentially with the distance $|z_{A_1}- z_{A_2}|$ between the two atoms. Thus, we can adjust the distance between the two atoms in the experiment such that $F(z_{A_1}, z_{A_2})/\omega_A=\varepsilon-s$, and $\alpha$ in our model is detemined by $g^2_c/2$.

The matrix elements of $\rho_{A_1}$ read respectively
\begin{equation}\label{Eq13}
\begin{split}
\rho_{A_1}^{++}(t)&\equiv\bra{+}\rho_{A_1}(t)\ket{+}\\
&=\frac{1}{2}|x_{A_2}|^2\left\{1-\cos[2\alpha(\varepsilon-s)t]\right\},\\
\rho_{A_1}^{+-}(t)&\equiv\bra{+}\rho_{A_1}(t)\ket{-}\\
&=ix_{A_2}y_{A_2}^*e^{-i\omega_A t}\sin\left[\alpha(\varepsilon-s)t\right].
\end{split}
\end{equation}
Here, we assume that $\varepsilon$ is subject to a Gaussian distribution with mean zero and the ensemble-averaged state $\langle\rho(t)\rangle$ defined in Eq.~(\ref{Eq3}) can be written as
\begin{align}\label{Eq14}
\langle\rho(t)\rangle_{++}\equiv&\overline{\rho_{A_1}^{++}(t)}\nonumber\\
=&\frac{1}{2}|x_{A_2}|^2\left[1-\cos(2\alpha st)e^{-2\alpha^2\overline{\varepsilon^2}t^2}\right],\nonumber\\
\langle\rho(t)\rangle_{+-}\equiv&\overline{\rho_{A_1}^{+-}(t)}\\
=&ix_{A_2}y_{A_2}^*e^{-i\omega_At}\sin(\alpha s t)e^{-\alpha^2\overline{\varepsilon^2}t^2/2}.\nonumber
\end{align}
Here, we have utilized the moment identity of a Gaussian distribution with mean zero, i.e.,  $\overline{\varepsilon^{2n}}=(\overline{\varepsilon^2})^n(2n)!/(2^nn!)$ and $\overline{\varepsilon^{2n-1}}=0$ ($n=1,2,\cdots$) \cite{Goodman2015}.

This result can be considered as the thermalization of the working qubit ${A_1}$ in a thermal bath, which is in a thermal equilibrium at temperature $T$. When the system A reaches the thermal equilibrium, its probability at the excited state is $P_+=e^{-\beta\Delta}(1+e^{-\beta\Delta})^{-1}$ \cite{Breuer2002}, with $\beta=1/k_{B}T$, and $k_B$ being the Boltzmann constant, where $\Delta$ is the energy-level difference between the two levels. The dissipation rate $\Gamma(t)=2\alpha^2\overline{\varepsilon^2}t$ is linear with respect to time $t$, which can be utilized to improve the quantum metrology and thus achieve Zeno limit \cite{Chin2012PRL,Matsuzaki2011PRA,Long2022PRL}. To simulate this steady-state distribution at arbitrary temperature $T$, we can effectively tune the initial state of the auxiliary qubit ${A_2}$, i.e., $x_{A_2}$ and $y_{A_2}$, to fulfill that $P_+=\overline{\rho_{A_1}^{++}(t\rightarrow \infty)}$. Notice that for any temperature $T$, this formula can always be satisfied because $0\leq x_{A_2} \leq1$, and thus $0\leq\overline{\rho_{A_1}^{++}(t\rightarrow \infty)}<1/2$.

\begin{figure}[!tb]
	\centering
	\includegraphics[width=8.4cm]{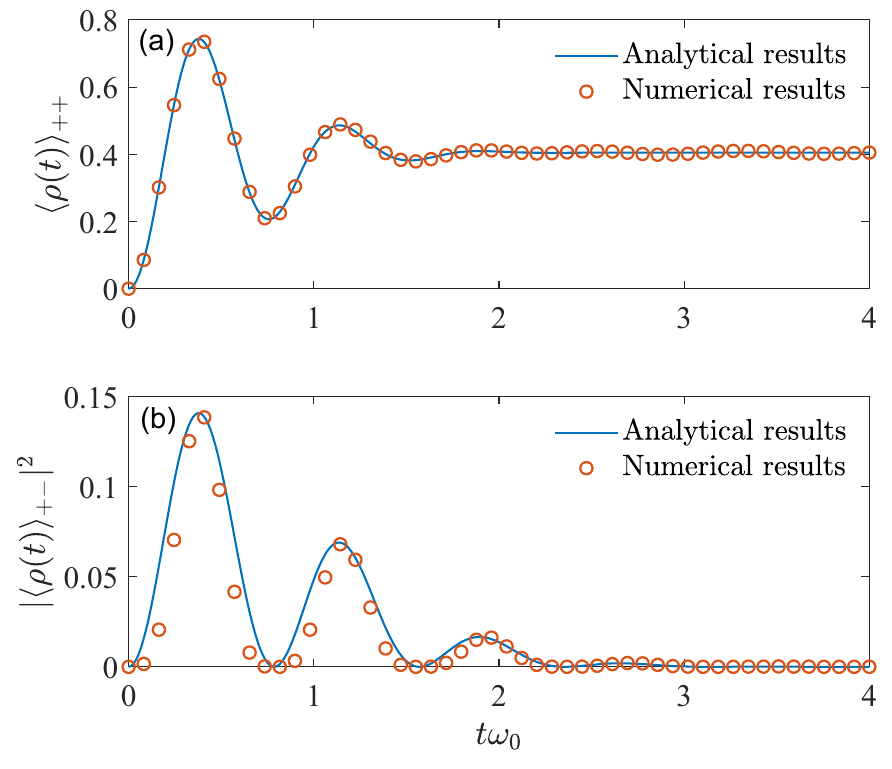}
	\caption{ Comparison of analytical and numerical results for the longitudinal relaxation of a single qubit. In the numerical calculation, we select $N=5000$ random samples $\left\{\varepsilon\right\}$ of Gaussian distribution with variance 0.6 and expectation 0. The quantum dynamics of (a) the excited-state population $\langle\rho(t)\rangle_{++}$, (b) the modular square of the coherence $|\langle\rho(t)\rangle_{+-}|^2$, when $s=4$, $x_{A_2}=0.9$, $\overline{\varepsilon^2}=0.6$, and $\alpha=1$.} 	\label{figure2}
\end{figure}

Here, as a demonstration, we simulate a process of a single qubit, initialized in the ground state, relaxation to the equilibrium state at a finite temperature $T$ through interaction with the heat reservoir. In this simulation, this process can be controlled by the initial state of the auxiliary qubit, the properties of the noise characterized by $\overline{\varepsilon^2}$ and the coupling strength characterized by $\alpha$. The behaviors of the longitudinal relaxation against the parameters, i.e., $x_{A_2}$, $\overline{\varepsilon^2}$ and $\alpha$, are plotted in Fig.~\ref{figure1}. In Fig.~\ref{figure1}(a), we leave $\overline{\varepsilon^2}$ and $\alpha$ unchanged and only vary $x_{A_2}$. We find that $x_B$ does not change the relaxation time but the steady-state population. In contrast, we can observe in Fig.~\ref{figure1}(b)(c), the relaxation time will decrease with the increase of the noise variance $\overline{\varepsilon^2}$ and the coupling strength $\alpha$, which do not change the steady-state population. The above observations are reasonable since the relaxation rate is $\Gamma(t)=2\alpha^2\overline{\varepsilon^2}t$ according to Eq.~(\ref{Eq13}). This relaxation process is not Markovian because the relaxation rate is time-dependent. We remark that the relaxation process can be Markovian if the bath-engineering technique is utlizied, i.e., the interaction strength between the working qubit and the auxiliary qubit is temporally tuned \cite{Soare2014PRA,Long2022PRL}. However, it is beyond the scope of the present investigation.


In order to verify our numerical simulation, the analytical and numerical results of full elements of the density matrix are compared in Fig.~\ref{figure2}. We find that when $s\ne0$, both the longitudinal and transverse relaxation behaviors demonstrate an oscillatory decay, but the decay of the transverse relaxation is slower. As shown in Fig.~\ref{figure2}(b), the interaction between the subsystem ${A_1}$ and the auxiliary qubit ${A_2}$ will induce the coherence between the ground state and the excited state, since the auxiliary qubit is initially in a superposition. However, when the steady state is reached, the coherence of the subsystem ${A_1}$ disappears and thus becomes a mixed state. Moreover, since the numerical results agree with the analytical results, our numerical simulations are reliable. To summarize, in this section, we utilize an auxiliary qubit to effectively simulate the longitudinal relaxation process of a single qubit at arbitrary temperature. It is found that the initial state $x_{A_2}$, frequency variance $\overline{\varepsilon^2}$ of the auxiliary qubit, and the interaction strength between the working qubit and the auxiliary qubit together determine the relaxation time and steady-state population of the longitudinal relaxation.

\section{Finite-Time Disentanglement}
\label{sec:disentanglement}

\begin{figure}[!tb]
	\centering
	\includegraphics[width=8.4cm]{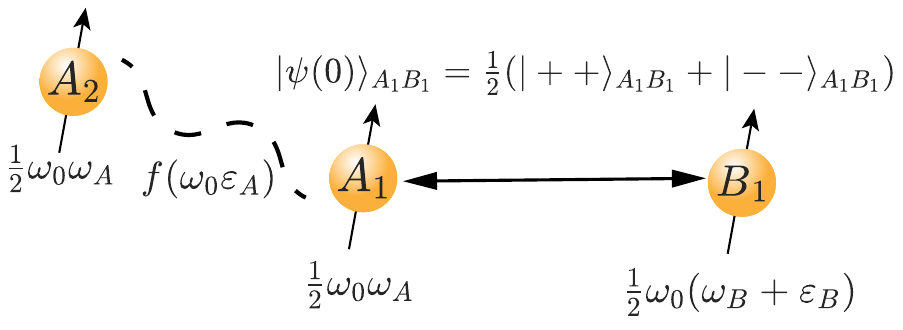}
	\caption{Schematic illustration of simulating sudden death of entanglement in a two-qubit system. Qubit $A_1$ and $B_1$ are initialized in the maximum-entangled state and have no interaction with each other. The random energy level-spacing characterized by $\varepsilon_B$ is used to simulate the transverse relaxation of $B_1$. We simulate the longitudinal noise of $A_1$ through the interaction between $A_1$ and the auxiliary qubit $A_2$.} 	\label{fig:TwoQubitSketch}
\end{figure}

In this section, we simulate the quantum dynamics of two-qubit disentanglement and investigate the effects of longitudinal and transverse relaxation on the disentanglement behavior. The system includes two working qubits, i.e., qubit $A_1$ and $B_1$, where the former interacts with an auxiliary qubit, i.e., qubit $A_2$, as schematically demonstrated in Fig.~\ref{fig:TwoQubitSketch}. Thus, the total Hamiltonian can be written in two parts as $\hat{H}_\varepsilon=\hat{H}_\varepsilon^A+\hat{H}_\varepsilon^B$, where
{\begin{align}\label{Eq15}
\hat{H}_\varepsilon^A=&\frac{\omega_0}{2}(\omega_A\sigma_{z}^{A_1}+\omega_A\sigma_{z}^{A_2})+f(\omega_0\varepsilon_A)\sigma_{+}^{A_1}\sigma_{-}^{A_2}+\textrm{h.c.},\nonumber\\
\hat{H}_\varepsilon^B=&\frac{\omega_0}{2}(\omega_B\sigma_{z}^{B_1}+\varepsilon_B\sigma_{z}^{B_1}).
\end{align}}
The composite system composed of $A_1$ and $B_1$ is initialized in the maximum-entangled state, i.e., $\ket{\psi(0)}_{A_1B_1}=(\ket{++}_{A_1B_1}+\ket{--}_{A_1B_1})/\sqrt{2}$. We let $A_1$ interact with an auxiliary qubit $A_2$ to mimic the longitudinal relaxation, as depicted in Sec.~\ref{sec:SingleQubit}, where $f(\varepsilon_A)$ is the coupling strength between $A_1$ and $A_2$. For qubit $B_1$, we apply a random energy-level spacing described by $\varepsilon_B$ to simulate the transverse relaxation \cite{Zhang2021FP,Wang2018NPJQI,Kropf2016PRX}.

\begin{figure}[!tb]
	\centering
	\includegraphics[width=8.4cm]{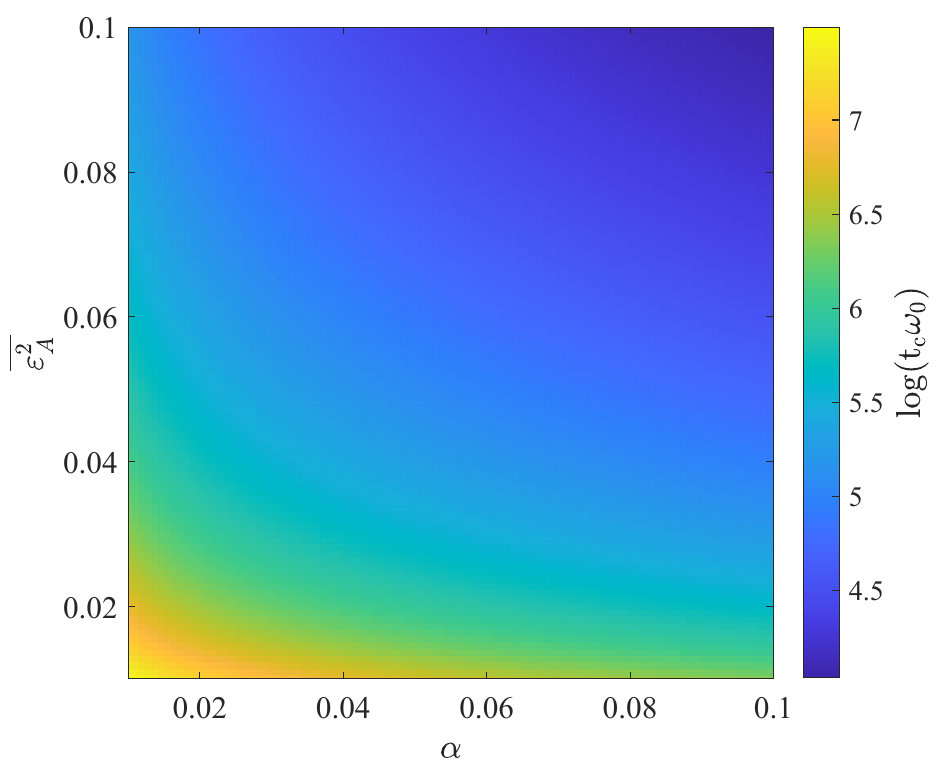}
	\caption{The critical disentanglement time $t_c$ against $\alpha$ and $\overline{\varepsilon_A^2}$ for $x=0.2$, $\overline{\varepsilon_B^2}=0$ and $s=0$. } 	\label{figure3}
\end{figure}

Before the ensemble average, we first of all solve the quantum dynamics of each realization. The initial state of the three qubits reads
\begin{align}\label{Eq16}
\rho(0)=x\ket{\psi_1(0)}\bra{\psi_1(0)}+y\ket{\psi_2(0)}\bra{\psi_2(0)},
\end{align}
where
\begin{equation}\label{Eq17}
\begin{split}
\ket{\psi_1(0)}&=\frac{1}{\sqrt{2}}\ket{+}_{A_2}\otimes(\ket{++}_{A_1B_1}+\ket{--}_{A_1B_1}),\\
\ket{\psi_2(0)}&=\frac{1}{\sqrt{2}}\ket{-}_{A_2}\otimes(\ket{++}_{A_1B_1}+\ket{--}_{A_1B_1}).
\end{split}
\end{equation}
As in the Sec.~\ref{sec:SingleQubit}, we define the coupling strength as $f(\varepsilon_A)=\alpha(\varepsilon_A-s)$. The ensemble-averaged state of the two working qubits, i.e., $\overline{\rho_{A_1B_1}(t)}$, can be written as
\begin{align}\label{Eq18}
	\overline{\rho_{A_1B_1}(t)}=\left(\begin{array}{cccc}
		\overline{a(t)} & 0 & 0 & 0 \\
		0 & \overline{b(t)} & \overline{z(t)} & 0 \\
		0 & \overline{z^*(t)} & \overline{c(t)} & 0 \\
		0 & 0 & 0 & \overline{d(t)}
	\end{array}\right).
\end{align}
Where we assume that $\varepsilon_A$ and $\varepsilon_B$ are subject to independent Gaussian distributions. The non-vanishing matrix elements of Eq.~(\ref{Eq18}) are explicitly given as
\begin{equation}\label{Eq19}
	\begin{split}
		\overline{a(t)}&=\frac{x}{4}\left[1-\cos(2\alpha s t)e^{-2\alpha^2\overline{\varepsilon_A^2}t^2}\right], \\
		\overline{b(t)}&=\frac{x}{2}+\frac{y}{4}\left[1+\cos(2\alpha s t)e^{-2\alpha^2\overline{\varepsilon_A^2}t^2}\right],\\
		\overline{c(t)}&=\frac{y}{2}+\frac{x}{4}\left[1+\cos(2\alpha s t)e^{-2\alpha^2\overline{\varepsilon_A^2}t^2}\right],\\
		\overline{z(t)}&=\frac{1}{2}e^{-\frac{1}{2}\overline{\varepsilon_B^2}t^2}e^{-\frac{i}{2}(\omega_A+\omega_B)t}e^{-\frac{1}{2}\overline{\varepsilon_A^2}\alpha^2t^2}\cos(\alpha s t),\\
		\overline{d(t)}&=\frac{y}{4}\left[1-\cos(2\alpha st)e^{-2\alpha^2\overline{\varepsilon_A^2}t^2}\right].
	\end{split}
\end{equation}
The detailed derivation of the above expressions appears in Appendix A.

\begin{figure}[!htb]
	\centering
	\includegraphics[width=8.4cm]{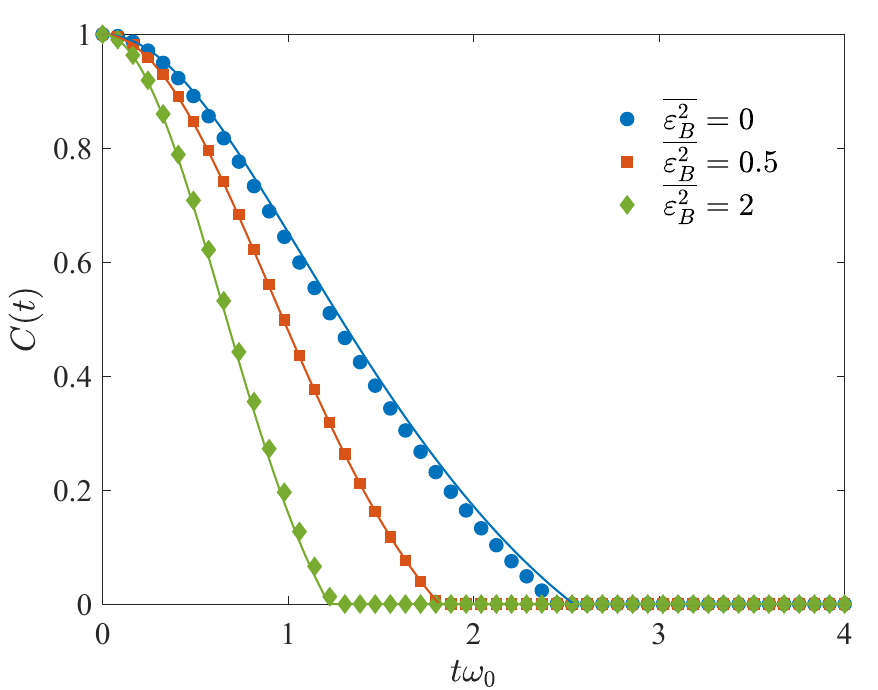}
	\caption{ The concurrence $C(t)$ as a function of time $t$ in the presence of both longitudinal and transverse relaxation for $\overline{\varepsilon_B^2}=0,~0.5,~2$ when $x=0.2$, $\overline{\varepsilon_A^2}=0.5$, $\alpha=1$, and $s=0$. Notice that the transverse relaxation is turned off when $\overline{\varepsilon_B^2}=0$. The analytic results are shown as solid lines. The dotted lines are generated by $N=300$ random samples with $\overline{\varepsilon_{A}}=\overline{\varepsilon_{B}}=0$.} 	\label{figure4}
\end{figure}

To investigate the disentanglement behavior of the composite system composed of $A_1$ and $B_1$ under both longitudinal and transverse relaxation, we utilize the concurrence \cite{Wootters1998PRL} to characterize the entanglement property $C(\overline{\rho_{A_1B_1}})=\text{max}(0,\sqrt{\kappa_1}-\sqrt{\kappa_2}-\sqrt{\kappa_3}-\sqrt{\kappa_4})$, where $\kappa_i$'s are the eigenvalues of the matrix $\mathcal{G}$ in decreasing order
\begin{align}\label{Eq20}
\mathcal{G} \equiv \overline{\rho_{A_1B_1}}\left(\sigma_y^A \otimes \sigma_y^B\right) \overline{\rho_{A_1B_1}}^*\left(\sigma_y^A \otimes \sigma_y^B\right),
\end{align}
where $\sigma_y^\alpha$ ($\alpha=A,B$) are the Pauli operators. When $C=1$, the two working qubits are in the maximum-entangled state, while $C=0$, they are disentangled with each other. Here, we can simplify the concurrence as
\begin{align}\label{Eq21}
C(\overline{\rho_{A_1B_1}})=2\text{max}\left\{0,|\overline{z}|-\sqrt{\overline{a}\overline{d}}\right\}.
\end{align}
At the beginning, $A_1$ and $B_1$ are initialized in the maximum-entangled state with $|\overline{z}|=1/2$ and $\sqrt{\overline{a}\overline{d}}=0$. In the following, we will show two categories of disentanglement in our simulation. In the first category, the entanglement tends to vanish only when the time approaches infinite. In the second category, the entanglement decays exactly to zero at a critical disentanglement time $t_c$ and remains zero thereafter.

\begin{figure}[!tb]
	\centering
	\includegraphics[width=8.4cm]{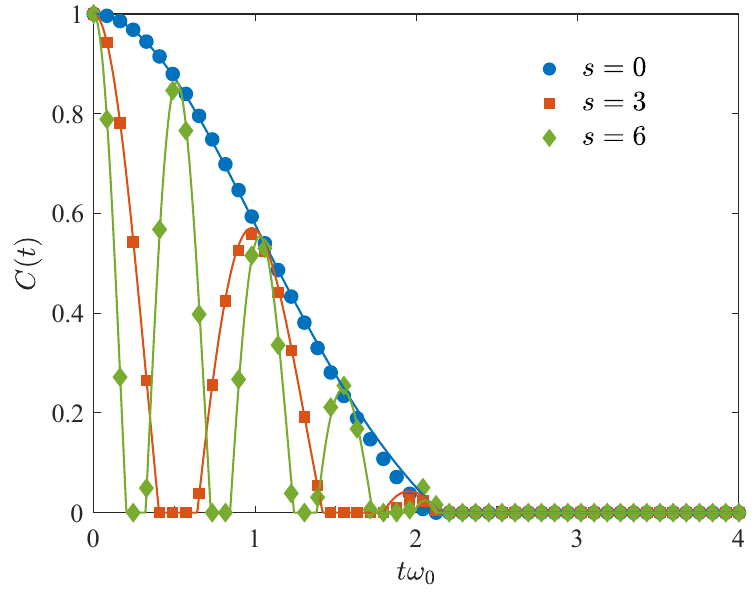}
	\caption{The concurrence $C(t)$ as a function of time $t$ in the presence of both longitudinal and transverse relaxation for $s=0,~3,~6$, when $x=0.2$, $\overline{\varepsilon_A^2}=0.5$, $\overline{\varepsilon_B^2}=0.2$, and $\alpha=1$. The analytic results are shown as solid lines. The dotted lines are generated by $N=300$ random samples with $\overline{\varepsilon_{A}}=\overline{\varepsilon_{B}}=0$.} 	\label{figure5}
\end{figure}

We first investigate the influence of the longitudinal relaxation on the entanglement properties when there is no transverse relaxation, i.e., $\overline{\varepsilon_B^2}=0$, in the system. In this case, the disentanglement behavior of the system is dominated by the coupling strength $\alpha$ and the noise fluctuation $\overline{\varepsilon_A^2}$. Obviously, when $\alpha=0$, i.e., the subsystem $A_1$ does not interact with the auxiliary qubit $A_2$, the system will always be in the maximum-entangled state. When $\alpha>0$, the non-vanishing noise fluctuation $\overline{\varepsilon_A^2}$ will determine whether the entanglement of the system can disappear at a finite time. When $\alpha>0$ and $\overline{\varepsilon_A^2}=0$, i.e., the longitudinal relaxation is turned off, $|\overline{z}|$ and $\sqrt{\overline{a}\overline{d}}$ can be written as
\begin{equation}\label{Eq22}
\begin{split}
|\overline{z}|&=\frac{1}{2}|\cos(\alpha st)|,\\
\sqrt{\overline{a}\overline{d}}&=\frac{\sqrt{xy}}{4}[1-\cos(2\alpha st)].
\end{split}
\end{equation}
such that $|\overline{z}|=1/2$ and $\sqrt{\overline{a}\overline{d}}=0$, and thus the entanglement of the system will not disappear persistently. However, if we turn on the longitudinal relaxation, i.e., $\alpha>0$ and $\overline{\varepsilon_A^2}>0$, $|\overline{z}|$ tends to zero and $\sqrt{\overline{a}\overline{d}}=\sqrt{xy}/4$ when the time approaches infinity. Thus, as long as $xy>0$, there exists a finite $t_c$, making the entanglement disappear after time $t_c$. The critical disentanglement time $t_c$ against $\overline{\varepsilon_A^2}$ and $\alpha$ without transverse relaxation is shown in Fig.~\ref{figure3}. We find that when $\alpha>0$ and $\overline{\varepsilon_A^2}>0$, the entanglement will disappear at a finite time and $t_c$ decays monotonically and rapidly as $\alpha$ and $\overline{\varepsilon_A^2}$ increase.

Now we consider the case when there is only transverse relaxation, i.e., $\overline{\varepsilon_B^2}>0$ and $\overline{\varepsilon_A^2}=0$. The evolution of $|\overline{z}|$ and $\sqrt{\overline{a}\overline{d}}$ can be written as
\begin{equation}\label{Eq23}
\begin{split}
|\overline{z}|&=\frac{1}{2}e^{-\frac{1}{2}\overline{\varepsilon_B^2}t^2}|\cos(\alpha st)|,\\
\sqrt{\overline{a}\overline{d}}&=\frac{\sqrt{xy}}{4}[1-\cos(2\alpha st)].
\end{split}
\end{equation}
Obviously, $|\overline{z}|$ tends to zero when the time approaches infinite. No matter how long the time passes, there always exist $2\alpha st=2n\pi$ with $n\in\mathcal{Z}$ so that $\sqrt{\overline{a}\overline{d}}$ vanishes. Thus, the concurrence will not constantly remain zero after a finite time but will go to zero at infinite time.

We can conclude that when the longitudinal relaxation exists, i.e., $\alpha>0$ and $\overline{\varepsilon_A^2}>0$, entanglement will disappear at a finite time. The larger the transverse noise fluctuates, the faster the entanglement disappears. And the oscillation frequency of $A_1$ hardly affects the time of disentanglement. The dynamics of concurrence against $\overline{\varepsilon_B^2}$ and $s$ in this case are shown in Figs.~\ref{figure4} and~\ref{figure5}, respectively. From Fig.~\ref{figure4}, we find that even if $\overline{\varepsilon_B^2}=0$, the concurrence will remain zero after the critical disentanglement time and decay faster with the increase of $\overline{\varepsilon_B^2}$. From Fig.~\ref{figure5}, we find that when the noise fluctuation and interaction strength are kept constant, the higher the frequency of subsystem $A_1$ is, the faster the entanglement $C$ oscillates, but the critical disentanglement time is almost the same. When the system is immune to the longitudinal relaxation, i.e., $\overline{\varepsilon_A^2}=0$ or $\alpha=0$, the entanglement will not die at a finite time but will disappear at infinite time due to the transverse relaxation.


\section{conclusion}
\label{sec:conclusion}

We have utilized the Hamiltonian-ensemble approach assisted by an auxiliary qubit to simulate the longitudinal relaxation of a single qubit in open quantum system. Concretely, the auxiliary qubits interacting with the working qubit are used to simulate the environmental effects. The theoretical results show that the simulated dynamics of the working qubit can be described by a real thermalization process. We can simulate the equilibrium-state distribution at any temperature, which is determined by the initial state of the auxiliary qubit, noise fluctuation and interaction strength. Furthermore, we simulate the dynamics of two-qubit entanglement initialized in the maximum-entangled state. We let the first qubit interact with the auxiliary qubit and relax longitudinally, and let the second qubit relax transversely. We find that if there is not longitudinal relaxation on the first qubit but transverse relaxation on the second qubit, the entanglement of the system will disappear when the time approaches infinity. However, if the longitudinal relaxation exists, no matter whether the transverse relaxation is present or not, the entanglement of the two-qubit will disappear after a finite time. And the larger the noise fluctuation is, the faster the entanglement will decay.

\begin{acknowledgments}
Q.A. thanks the financial support from Beijing Natural Science Foundation under Grant No.~1202017 and the National Natural Science Foundation of China under Grant Nos.~11674033,~11505007, and Beijing Normal University under Grant No.~2022129. H.B.W. thanks the financial support from National Natural Science Foundation of China under Grant No.~61675028 and National Natural Science Foundation of China under Grant No.~12274037.
\end{acknowledgments}

\appendix
\section{DERIVATION OF THE ENSEMBLE AVERAGED STATE OF THE TWO WORKING QUBITS}
The initial state of the three qubits for each realization reads
\begin{align}\label{EqA1}
	\rho(0)=x\ket{\psi_1(0)}\bra{\psi_1(0)}+y\ket{\psi_2(0)}\bra{\psi_2(0)},
\end{align}
where
\begin{equation}\label{EqA2}
	\begin{split}
		\ket{\psi_1(0)}&=\frac{1}{\sqrt{2}}\ket{+}_{A_2}\otimes(\ket{++}_{A_1B_1}+\ket{--}_{A_1B_1}),\\
		\ket{\psi_2(0)}&=\frac{1}{\sqrt{2}}\ket{-}_{A_2}\otimes(\ket{++}_{A_1B_1}+\ket{--}_{A_1B_1}).
	\end{split}
\end{equation}
In other words, the two working qubits $A_1$ and $B_1$ are in a maximum-entangled state, while the auxiliary qubit $A_2$ is in a mixed state. At time $t$, $\ket{\psi_1(0)}$ evolves into the state
\begin{eqnarray}\label{EqA3}
	\ket{\psi_1(t)}&=&\eta_1\ket{+++}_{A_2A_1B_1}+\eta_2\ket{-+-}_{A_2A_1B_1}\nonumber\\
	&&+\eta_3\ket{+--}_{A_2A_1B_1},
\end{eqnarray}
where $\eta_1=\exp\left[-i(\omega_A+\varepsilon_A+\omega_B+\varepsilon_B)t/2\right]/\sqrt{2}$, because $\ket{+++}_{A_2A_1B_1}$ is the eigenstate of $\hat{H}_\varepsilon$. In the invariant subspace spanned by the basis $\{\ket{-+-}_{A_2A_1B_1},\ket{+--}_{A_2A_1B_1}\}$, the effective Hamiltonian can be simplified as
\begin{align}\label{EqA4}
	\hat{H}_\varepsilon=-\frac{1}{2}(\omega_B+\varepsilon_B)I+f(\varepsilon_A)\sigma_x.
\end{align}
And thus the evolution operator $\hat{U}_\varepsilon=\exp(-i\hat{H}_\varepsilon t)$ reads
\begin{align}\label{EqA5}
	\hat{U}_\varepsilon&=[\cos(f(\varepsilon_A)t)-i\sin(f(\varepsilon_A)t)\sigma_x]e^{\frac{i}{2}(\varepsilon_B+\omega_B)t}.
\end{align}
Since $\left[\eta_2, \eta_3\right]^T=\hat{U}_\varepsilon\left[0,1/\sqrt{2}\right]^T$, the three coefficients of $\ket{\psi_1(t)}$ are explicitly given as
\begin{equation}\label{EqA6}
	\begin{split}
		\eta_1\!\!&=\!\!\frac{1}{\sqrt{2}}e^{-\frac{i}{2}(2\omega_A+\omega_B+\varepsilon_B)t},\\
		\eta_2\!\!&=\!\!\frac{-i}{\sqrt{2}}e^{\frac{i}{2}(\varepsilon_B+\omega_B)t}\sin[f(\varepsilon_A)t],\\
		\eta_3\!\!&=\!\!\frac{1}{\sqrt{2}}e^{\frac{i}{2}(\varepsilon_B+\omega_B)t}\cos[f(\varepsilon_A)t].
	\end{split}
\end{equation}

Suppose $\ket{\psi_2(t)}$ can be expanded as
\begin{eqnarray}\label{EqA7}
		\ket{\psi_2(t)}&=&\xi_1\ket{---}_{A_2A_1B_1}+\xi_2\ket{+-+}_{A_2A_1B_1}\nonumber\\
		&&+\xi_3\ket{-++}_{A_2A_1B_1}.
\end{eqnarray}
Following the above steps, we can obtain
\begin{equation}\label{EqA8}
	\begin{split}
		\xi_1\!\!&=\!\!\frac{1}{\sqrt{2}}e^{\frac{i}{2}(2\omega_A+\omega_B+\varepsilon_B)t},\\
		\xi_2\!\!&=\!\!\frac{-i}{\sqrt{2}}e^{-\frac{i}{2}(\varepsilon_B+\omega_B)t}\sin[f(\varepsilon_A)t],\\
		\xi_3\!\!&=\!\!\frac{1}{\sqrt{2}}e^{-\frac{i}{2}(\varepsilon_B+\omega_B)t}\cos[f(\varepsilon_A)t].
	\end{split}
\end{equation}
Because we have solved the quantum dynamics of the three qubits, by tracing over the auxiliary qubit $A_2$, we can obtain the reduced density matrix of the two working qubits $\rho_{A_1B_1}(t)=\operatorname{Tr}_{A_2}\rho(t)$ in the basis $\{\ket{+-}_{A_1B_1},\ket{++}_{A_1B_1},\ket{--}_{A_1B_1},\ket{-+}_{A_1B_1}\}$
as
\begin{align}\label{EqA9}
	\rho_{A_1B_1}(t)=\left(\begin{array}{cccc}
		a(t) & 0 & 0 & 0 \\
		0 & b(t) & z(t) & 0 \\
		0 & z^*(t) & c(t) & 0 \\
		0 & 0 & 0 & d(t)
	\end{array}\right).
\end{align}
As in the Sec.~\ref{sec:SingleQubit}, we define the coupling strength as $f(\varepsilon_A)=\alpha(\varepsilon_A-s)$, where $\alpha\geq0$. The non-vanishing matrix elements are given by
\begin{align}\label{EqA10}
		a(t)=&\frac{x}{2}\gamma(t)^2, \nonumber\\
		b(t)=&\frac{x}{2}+\frac{y}{2}\left[1-\gamma(t)^2\right], \nonumber\\
		c(t)=&\frac{y}{2}+\frac{x}{2}\left[1-\gamma(t)^2\right], \\
		z(t)=&\frac{1}{2}\zeta(t)
		\cos[\alpha(\varepsilon_A-s)t],\nonumber\\
		d(t)=&\frac{y}{2}\gamma(t)^2, \nonumber
\end{align}
where $\zeta(t)=\exp\left[-i(\omega_A+\omega_B+\varepsilon_B)t\right]$, $\gamma(t)=\sin[\alpha(\varepsilon_A-s)t]$. We assume that $\varepsilon_A$ and $\varepsilon_B$ are subject to independent Gaussian distributions. After the ensemble average, the non-vanishing matrix elements of Eq.~(\ref{EqA9}) are explicitly given as
\begin{equation}\label{EqA11}
		\begin{split}
			\overline{a(t)}&=\frac{x}{4}\left[1-\cos(2\alpha s t)e^{-2\alpha^2\overline{\varepsilon_A^2}t^2}\right], \\
			\overline{b(t)}&=\frac{x}{2}+\frac{y}{4}\left[1+\cos(2\alpha s t)e^{-2\alpha^2\overline{\varepsilon_A^2}t^2}\right],\\
			\overline{c(t)}&=\frac{y}{2}+\frac{x}{4}\left[1+\cos(2\alpha s t)e^{-2\alpha^2\overline{\varepsilon_A^2}t^2}\right],\\
			\overline{z(t)}&=\frac{1}{2}e^{-\frac{1}{2}\overline{\varepsilon_B^2}t^2}e^{-\frac{i}{2}(\omega_A+\omega_B)t}e^{-\frac{1}{2}\overline{\varepsilon_A^2}\alpha^2t^2}\cos(\alpha s t),\\
			\overline{d(t)}&=\frac{y}{4}\left[1-\cos(2\alpha st)e^{-2\alpha^2\overline{\varepsilon_A^2}t^2}\right].
		\end{split}
\end{equation}

\providecommand{\noopsort}[1]{}\providecommand{\singleletter}[1]{#1}%
%


\end{document}